\documentclass[doublecol]{epl2}

\title{Numerical study of Anderson localization of terahertz waves\\ in disordered waveguides}
\shorttitle{Anderson localization of terahertz waves} 

\author{C. P. Lapointe\inst{1} \and P. Zakharov\inst{1} \and F. Enderli\inst{2} \and T. Feurer\inst{2} \and S. E. Skipetrov\inst{3} \and F. Scheffold\inst{1}}
\shortauthor{C. P. Lapointe \etal}

\institute{
  \inst{1} Physics Department, University of Fribourg, Chemin du Mus\'{e}e 3, 1700 Fribourg, Switzerland\\
  \inst{2} Institute of Applied Physics, University of Bern, Siedlerstrasse 5, 3012 Bern, Switzerland\\
   \inst{3} Universit\'{e} Grenoble 1/CNRS, LPMMC UMR 5493, B.P. 166, 38042 Grenoble, France\\
}
\pacs{42.25.Dd}{Wave propagation in random media}
\pacs{42.25.Bs}{Wave propagation, transmission and absorption}
\pacs{71.55.Jv}{Disordered structures; amorphous and glassy solids}

\abstract{We present a numerical study of electromagnetic wave transport in disordered quasi-one-dimensional waveguides at terahertz frequencies. Finite element method calculations of terahertz wave propagation within LiNbO$_{3}$ waveguides with randomly arranged air-filled circular scatterers exhibit an onset of Anderson localization at experimentally accessible length scales. Results for the average transmission as a function of waveguide length and scatterer density demonstrate a clear crossover from diffusive to localized transport regime. In addition, we find that transmission fluctuations grow dramatically when crossing into the localized regime. Our numerical results are in good quantitative agreement with theory over a wide range of experimentally accessible parameters both in the diffusive and localized regime opening the path towards experimental observation of terahertz wave localization.}

\begin{document}

\maketitle

\section{Introduction}

More than fifty years after Philip Anderson suggested the localization of de Broglie electron waves \cite{Anderson58} its analogous manifestation involving classical electromagnetic (EM) waves is still under discussion \cite{LagendijkPhysTod09,Wiersma13NP}. In particular, several reported observations of localization of light in three dimension have been questioned \cite{WiersmaNature97,Maret13}. A recent theoretical study even suggests the absence of light localization in a random three-dimensional (3D) ensemble of point scatterers \cite{Skipetrov2013noloc}. The situation appears to be much clearer in constrained geometries having a finite number $N$ of propagating EM modes at a given frequency $f$. In particular, the physics of wave transport in disordered waveguides of length $L$ much exceeding their width $w$ is currently well established \cite{Mello91,MStoytchevPRL1997,AAChabanovNature2000,AZGenackJPhysAMathGen2005,vanRossumRevModPhys1999,CWBeenakerRevModPhys1997,MirlinPhysRep2000,Froufe2002}. The latter geometry is referred to as quasi-one-dimensional (quasi-1D) when $w$ is smaller or of the order of the transport mean free path $l^*$ due to disorder because transport equations for ensemble-averaged quantities (such as, e.g., the average intensity) turn out to involve only one spatial dimension. In contrast to the case of fully 3D systems, Anderson localization should always take place provided that the dimensionless conductance $g_0\sim N l^{\ast}/L$ is less than one. A complete set of data on Anderson localization in quasi-1D disordered systems has been presented by the group of Azriel Genack  about a decade ago \cite{MStoytchevPRL1997,AAChabanovNature2000,AZGenackJPhysAMathGen2005}. These experiments concerned microwaves propagating in hollow copper tubes filled with random assemblies of  spheres having sizes of the order of the wavelength of the propagating radiation. More recent experiments on light propagation in silicon waveguides further support the localization scenario in quasi-1D waveguide geometries \cite{Cao13}.

In this article, we present a numerical study of EM wave transport in disordered quasi-1D waveguides at terahertz frequencies. The interest of studying Anderson localization of THz waves is twofold. First, the refractive index contrast between the scatterers and the background medium can be exceptionally high, of the order of 5 to 1 for common materials. Second, it has been shown that the electric field amplitude of THz waves can be imaged non-invasively \cite{TFeurerReviewTHz2007,NSStoyanovTHzWGs2003}, thus providing a unique tool to study the spatial distribution of the electric field in and close to the localized regime with high spatial resolution. We expect that coherent terahertz waves in ferroelectric crystals will provide a powerful system, outside of established fields such as optical-frequency EM waves \cite{WiersmaNature97,Maret13}, microwaves \cite{MStoytchevPRL1997,AAChabanovNature2000,AZGenackJPhysAMathGen2005}, acoustic waves \cite{AcousticLocalization2008} and matter waves \cite{Billy2008}, to quantitatively study Anderson localization effects. As a first step towards this goal, we present a numerical study of terahertz wave transport in quasi-1D disordered waveguides for which exact \cite{MirlinPhysRep2000} and approximate \cite{PaynePRB2010,Yamilov12} theoretical predictions are available for the the average transport coefficients near the onset of Anderson localization, as well as for the full distribution function of transmission fluctuations \cite{vanRossumRevModPhys1999,Nieuwen1995}. These quantities can be directly measured in future experiments. Moreover, we establish crucial parameters of the system, which maximize localization effects in structured LiNbO$_{3}$ waveguides in order to guide future experimental and theoretical studies. We note that our approach explicitly models the experimental situation in all relevant details, and is therefore distinctive from many other numerical studies in which either point scatterers \cite{PaynePRB2010,Yamilov12} or surface roughness \cite{SaenzSurfaceRoughnessPRL1998,SaenzSurfaceRoughLocalizationPRL1998,SaenzFiniteSizePRL2002} induce diffuse transport and localization.

\section{Phonon-polaritons in ferroelectric crystals}

Coupled electromagnetic-lattice vibrational waves, referred to as phonon-polaritons, can be generated within ferroelectric crystals through impulsive stimulated Raman scattering, or alternatively, through optical rectification by irradiating a crystal with a pulsed femtosecond laser \cite{TFeurer2003}. For the ferroelectric crystal LiNbO$_3$, at frequencies below 2 THz, phonon-polariton dispersion is light-like with propagation speeds of \textit{c}/\textit{n} where \textit{c} is the speed of light in vacuum and the refractive index is \textit{n} = 5.1 \cite{TFeurerReviewTHz2007}. Recently developed pump-probe techniques allow for the generation and imaging of both the time-dependent amplitude and phase of phonon-polaritonic fields \cite{TFeurerReviewTHz2007,NSStoyanovTHzWGs2003}. In addition to powerful imaging methods, scattering sites can be readily fabricated in the form of micrometer-scale cylindrical air-filled holes in a LiNbO$_{3}$ crystal using micro-machining methods such as laser ablation \cite{StoyanovNatMat2002,CAWerleyTHzStrucWGs2010, PPeierTHzStrucWGs2010} and optical lithography \cite{Benchabane2006}. The spatial resolution requirements for both machining structured waveguides and imaging propagating polaritonic waves in ferroelectric crystals are nearly two orders of magnitude less stringent than is the case for visible light. For instance, the wavelength of a phonon-polariton at a frequency of 1 THz in LiNbO$_{3}$ is roughly 60 $\mu$m. The large refractive index contrast between LiNbO$_{3}$ and air at terahertz frequencies induces strong scattering of phonon-polaritons from the holes, leading to a situation that is favorable for the observation of Anderson localization.

\section{Simulations of terahertz wave transport}

Since dispersion of phonon-polariton waves in LiNbO$_3$ is light-like, we simulate phonon-polaritonic transport through disordered quasi-1D waveguides by solving the Helmholtz equation for the electromagnetic field in the frequency domain using finite element methods in two dimensions. The waveguide is supposed to be rectangular, with the length $L$ much exceeding the width $w$. Perfectly reflecting boundary conditions are imposed at the outer edges of the waveguide. To eliminate back reflections at the ends of the waveguide, a 0.5 mm-long perfectly matched layer is used. It effectively absorbs all incident radiation over approximately one wavelength.

To tune the scattering strength in the system, calculations are performed for two scatterer radii $a = 25$ and 50 $\mu$m, and at three excitation frequencies $f = 0.5$, 1 and 1.5 THz. These parameters are readily accessible in experiments using LiNbO$_{3}$ crystals \cite{TFeurerReviewTHz2007}. We consider air-filled scatterers having refractive index $n_{\mathrm{sc}} = 1$ and fix the width of the waveguide to $w = 500$ $\mu$m. The presence of the scatterers lowers the effective refractive index of the waveguide $n_{\mathrm{eff}}$. Using Bruggeman effective medium theory \cite{BookSihvola}, we find that $n_{\mathrm{eff}}$ decreases from 5.1 at areal filling fraction of the scatterers $\phi = 0$ down to 4.93 and 4.09, at $\phi = 0.0193$ and 0.196, respectively, for $a = 50$ $\mu$m. Over a 0.5--1.5 THz frequency range, the number of transverse propagating modes $N = kw/\pi$ is therefore $N=7$ to 25 for an empty waveguide and $N = 6$ to 24 for the areal filling fractions considered here.

\begin{figure}[t]
\centerline{\includegraphics[width=0.9\columnwidth]{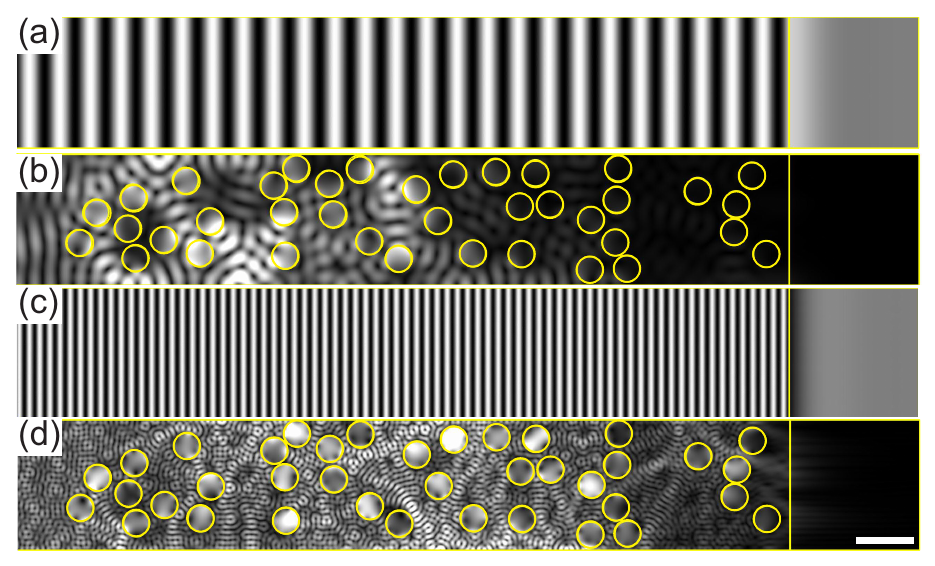}}
\caption{Grayscale plots of intensity inside the waveguide without scatterers and for $\phi = 0.196$ and $a = 50$ $\mu$m at frequencies $f = 0.5$ THz (a, b) and $f = 1.5$ THz (c, d). Open circles show the boundaries of the circular air-filled scatterers in the waveguide. Scale bar: 250 $\mu$m.} \label{Fig1}
\end{figure}

To calculate the average transmission, ensembles consisting of 100 arrangements of randomly positioned circular scatterers are generated at areal filling fractions up to $\phi \sim 0.2$. To generate a given arrangement, scatterers are successively added to the system using a random number generator with the constraint that no two scatterer centers lie closer than $2(a + 1)$ $\mu$m. The additional 1 $\mu$m is employed to avoid meshing problems during simulation runs. This method of placing circular scatterers results in a hard-sphere liquid-like pair correlation distribution, the multiple scattering properties of which are well studied both theoretically \cite{Kaplan1992} and experimentally \cite{Fraden1990,RojasOchoaPRE2002,RojasOchoaPRL2004}.

An adaptive, variable-size Delaunay triangular grid is created using commercially available finite element simulation software (COMSOL Multiphysics 4.3). The tradeoff between numerical accuracy and computation time dictates using an average grid point density of $430$ $\mu$m$^{-2}$ having maximum and minimum spacing of 10 and 4.4 $\mu$m, respectively. Results differ by less than 1\% after increasing the total number of grid points by a factor of 30. The input of the waveguide is excited with a monochromatic plane wave polarized transverse to the long axis of the waveguide and solutions to the two-dimensional Helmholtz equation for the in-plane electromagnetic field are found. For illustration purposes, we show in Fig.\ \ref{Fig1} grayscale plots of the wave intensity inside the waveguide at two different frequencies for randomly placed 50 $\mu$m radius scatterers at a filling fraction $\phi = 0.196$. Transmission coefficients are determined by calculating the power transmitted downstream from the scatterers, and then dividing the result by the power transmitted through the same waveguide without scatterers.

\section{Results}
Next we present the results or our a numerical study of terahertz wave transport in quasi-1D disordered waveguides and it's comparison to the exact \cite{MirlinPhysRep2000} and approximate \cite{PaynePRB2010,Yamilov12} theoretical predictions reported in the literature.

\subsection{Average transmission}
\begin{figure}[t]
\centerline{\includegraphics[width=0.9\columnwidth]{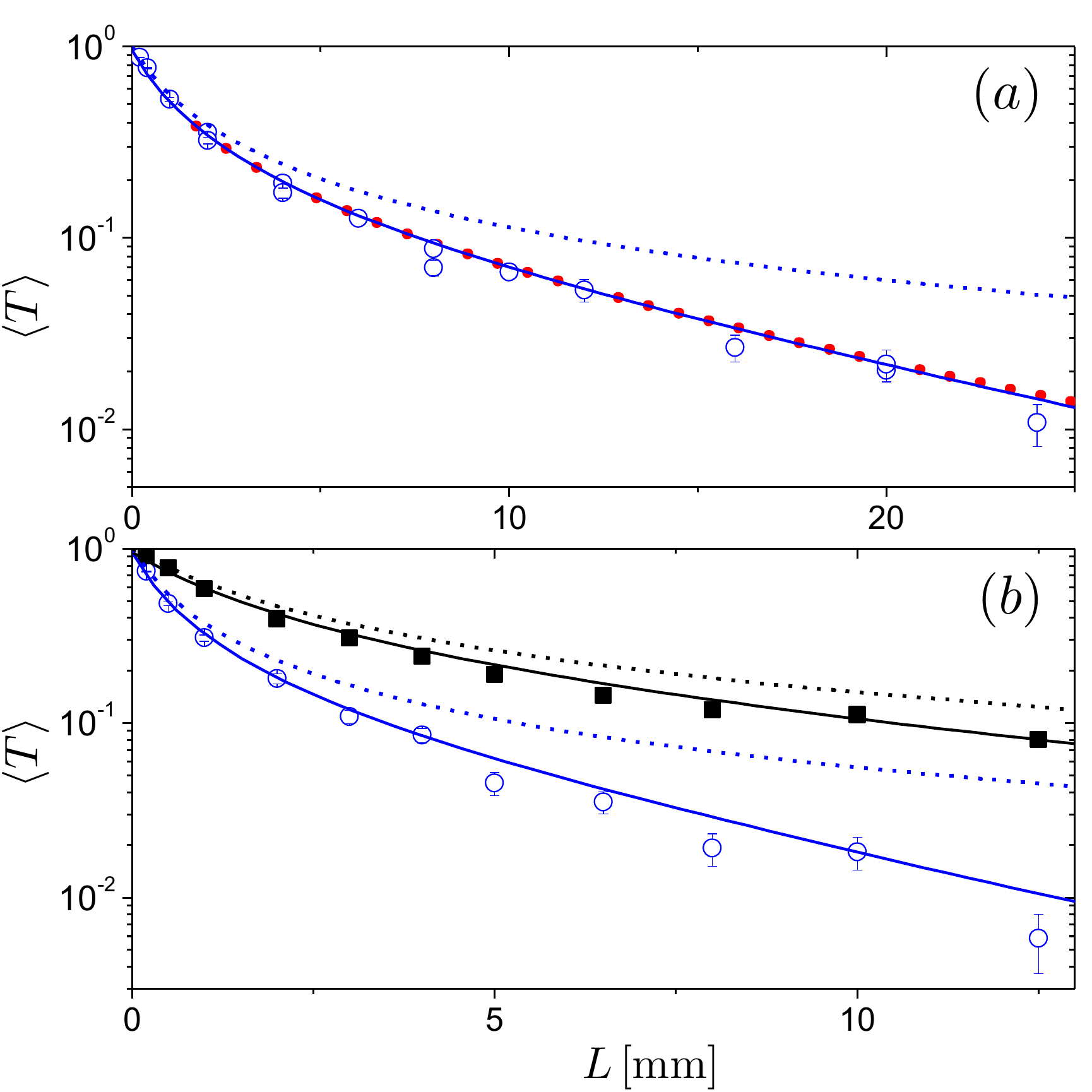}}
\caption{Average transmission through a LiNbO$_3$ disordered waveguide of width $w = 0.5$ mm as a function of length $L$ at $f = 0.5$ THz, $N = 7$ for (a) $\phi = 0.079$ and (b) $\phi = 0.157$, $n_{\mathrm{sc}} = 1$ (open circles) and $n_{\mathrm{sc}} = 2$ (squares). Solid lines are fits to Eq.\ (\ref{power series}) using $l^*$ as the only free parameter ($l^* = 0.81$, 1.13, 0.38 mm from top to bottom). Full circles in (a) show the prediction of the exact theory for $\langle g \rangle$ \cite{MirlinPhysRep2000} divided by $N$, with $l^* = 0.81$ mm. The localization length is $\xi = (\pi/2) N l^* \simeq 9$ mm.   Predictions for purely diffusive transport, $\langle T_0 \rangle = g_0/N$ with $g_0$ from Eq.\ (\ref{dimcond}), are shown by the dotted curves using the same values of $l^*$. Error bars indicate standard deviation of the mean due to statistical fluctuations of results from one realization of disorder to another.} \label{Fig2}
\end{figure}

Representative results for average transmission $\langle T \rangle$ as a function of waveguide length are shown in Fig.\ \ref{Fig2} for different areal densities of scatterers at the frequency $f = 0.5$ THz. To characterize the influence of scatterer density on the crossover from diffusive to localized transport, we calculated the average transmission for waveguides having various areal filling fractions of air-filled scatterers ranging from $\phi = 0.039$ to 0.196 for $a = 50$ $\mu$m and from $\phi = 0.049$ to 0.197 for $a = 25$ $\mu$m. For brevity, we focus our discussion on results for \textit{a} = 50 $\mu$m; similar results are obtained for all three frequencies and both scatterer sizes explored in this study. To keep the same level of accuracy throughout the theoretical discussion, we use simplified expressions for the dimensionless conductance $g_0 $ and the average diffuse transmission $\langle T_0 \rangle \simeq g_0/N$ taking into account internal reflections of waves at the entry and exit of the waveguide through the so-called extrapolation length $z_0 = (\pi/4) l^*$ \cite{vanRossumRevModPhys1999}:
\begin{equation}
g_0 = \left( \frac{L}{\xi} + \frac{1}{N} \right)^{-1},
\label{dimcond}
\end{equation}

where $\xi = (\pi/2) Nl^*$ is the localization length. For $L \to 0$, we obtain $g_0 = N$ and thus $\langle T_0 \rangle = 1$, whereas for  $L/\xi \gg 1/N$ we get $g_0 = \xi/L$. The relation $\langle T \rangle \simeq g/N$ between $\langle T \rangle$ and $g$ assumes that all $N$ channels are equivalent and that on average, the multiple scattering redistributes the incident radiation uniformly among $N$ outgoing transverse modes. We note that in the diffusive regime, it is possible to take interfacial effects into account more precisely (see, e.g., Ref.\ \cite{PaynePRB2010}), but the corrections due to localization effects, both for the average transmission
$\langle T \rangle$ and its probability distribution $P(T/\langle T \rangle)$ (see below), cannot be evaluated to the same level of accuracy.

Fits using the diffuse regime result --- Eq.\ (\ref{dimcond}) and $\langle T \rangle = \langle T_0 \rangle = g_0/N$ --- deviate appreciably from the data for waveguide lengths $L \ge \xi$, see Fig.\ \ref{Fig2}. The deviation from the classical diffusive transport indicates the presence of localized states and the concomitant onset of interference effects. Next we compare the numerical data to a result from the self-consistent theory of Anderson localization from which the average transmission $\langle T \rangle$ can be expressed as a power series in $1/g_0$ \cite{PaynePRB2010}:
\begin{equation}
\langle T \rangle  \simeq \frac{g_0}{N} \left[ 1 - \frac{1}{3 g_0} + \frac{1}{45g_0^2} + \frac{2}{945 g_0^3} + \ldots \right].
\label{power series}
\end{equation}
\begin{figure*}[t]
\centerline{\includegraphics[width=2\columnwidth]{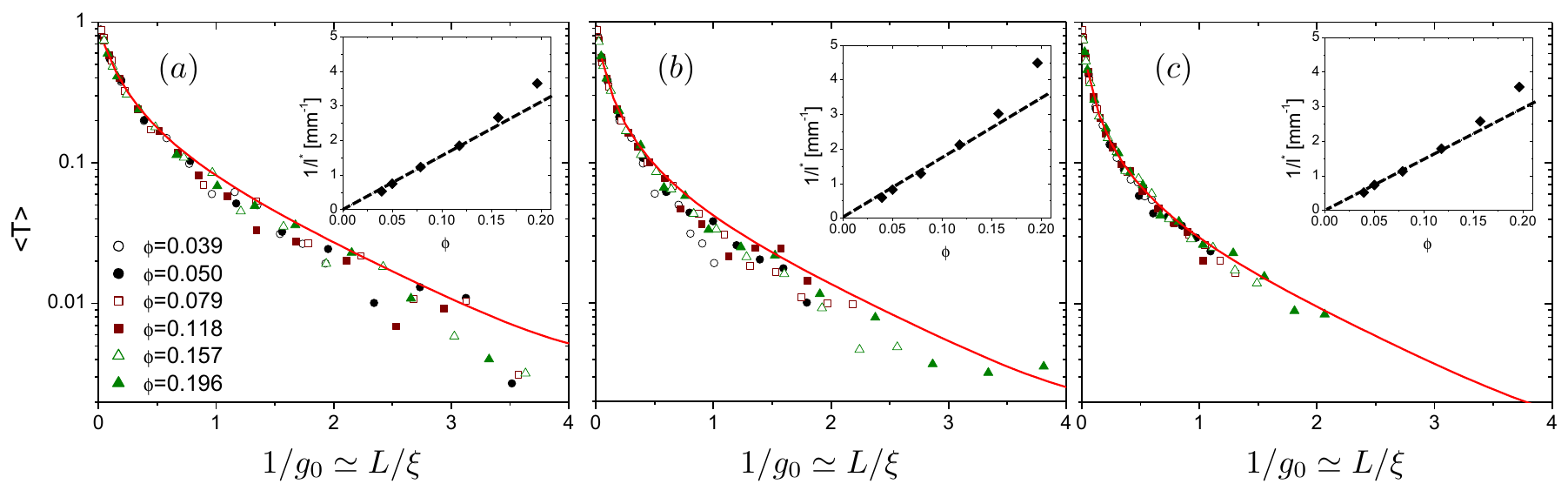}}
\caption{Average transmission $\langle T \rangle$ for all filling fractions from $\phi=0.039$ to 0.196 shown as a function of $L/\xi \simeq 1/g_0$ for three different frequencies $(a)$ f=0.5 THz, $(b)$ 1 THz, $(c)$ 1.5 THz and $n_{\mathrm{sc}} = 1$. Solid lines are predictions of Eq.\ (\ref{power series}), truncated after the fourth term, with $N = 7$, 15, 22 for frequencies $f = 0.5$, 1, 1.5 THz, respectively. Insets show the inverse transport mean free path $1/l^*$ versus the scatterer areal filling fraction. Dotted lines show least squares fits to the first 4 data points as discussed in the text.}
\label{Fig3}
\end{figure*}

As can be seen in Fig.\ \ref{Fig2}, fits to the numerical data using Eq.\ (\ref{power series}) with $l^*$ as the only fit parameter are excellent. Moreover, we illustrate in Fig.\ \ref{Fig2}(a) that the approximations made in the series expansion, Eq.\ (\ref{power series}) truncated after the fourth term, are small as long as $L/\xi < 3$ (in Fig. \ref{Fig2}(a) compare the solid line and full circles, the latter being obtained using exact results for $\langle g \rangle$ \cite{MirlinPhysRep2000}). This is, in fact, not surprising since Eq.\ (\ref{power series}) coincides with the series expansion of the exact result in Ref.\cite{MirlinPhysRep2000} and can also be obtained from the maximum-entropy model based on a transfer-matrix formulation \cite{Mello91}. Further evidence for the quantitative agreement of numerical data and theory is given in Fig.\ \ref{Fig3} where we show that for any given frequency, the data can be collapsed onto a master curve when plotted as a function of $1/g_0$, as predicted by Eq.\ (\ref{power series}).

The analysis of a number of areal filling fractions for different frequencies allows us to study the dependence of transport mean free path on scatterer density. We find that the inverse transport mean free path $1/l^*$ obtained from the fits increases linearly with $\phi$ for $\phi \le 0.12$, whereas at larger $\phi$, $1/l^*$ grows slightly faster. For arrangements of monodisperse scatterers having no spatial correlations, the transport mean free path can be expressed as $l^* = l/(1-\gamma)$, where $l = 1/(Q_{\mathrm{sc}}\phi)$ is the scattering mean free path and $\gamma$ is the scattering anisotropy parameter. A linear fit having no intercept to the first four points in the inset of Fig.\ \ref{Fig3}(b), and using $\gamma = 0.5$, yields $Q_{\mathrm{sc}} = 32$ mm$^{-1}$. This is not too far from the value obtained using exact Mie scattering theory for a long cylinder in an infinite plane \cite{BookBohrenHuffman}. Using the latter theory we find that as a function of frequency, $Q_{\mathrm{sc}}$ is peaked at $f = 1$ THz with a maximum of $Q_{\mathrm{sc}} \simeq 19.1$ mm$^{-1}$ and that the anisotropy parameter $\gamma$ is approximately equal to 0.5 over the frequency range from 0.5 to 1.5 THz. We note that we do not expect $Q_{\mathrm{sc}}$ of a cylindrical scatterer in the waveguide and in the homogenous infinite space to be identical. It will somewhat depend on the finite number of transverse modes accessible for the scattered waves \cite{Juanjo2001}. Moreover, there might be some effect due to the finite scatterer size as it becomes comparable to the width of the waveguide. In addition, at higher values of $\phi$, $1/l^*$ deviates slightly from the linear scaling  $1/l^* \propto \phi$ possibly because of short-range positional correlations, which are known for dense colloidal dispersions to either decrease \cite{Fraden1990, Kaplan1992} or increase \cite{RojasOchoaPRE2002,RojasOchoaPRL2004} $l^*$ relative to the uncorrelated system, depending upon the size, number density and refractive index of the scatterers. However, the study of the complex interplay between these competing effects and their influence on $l^*$ is beyond the scope of the present article.

\subsection{Fluctuations of transmission}

\begin{figure}[t]
\centerline{\includegraphics[width=8.8cm]{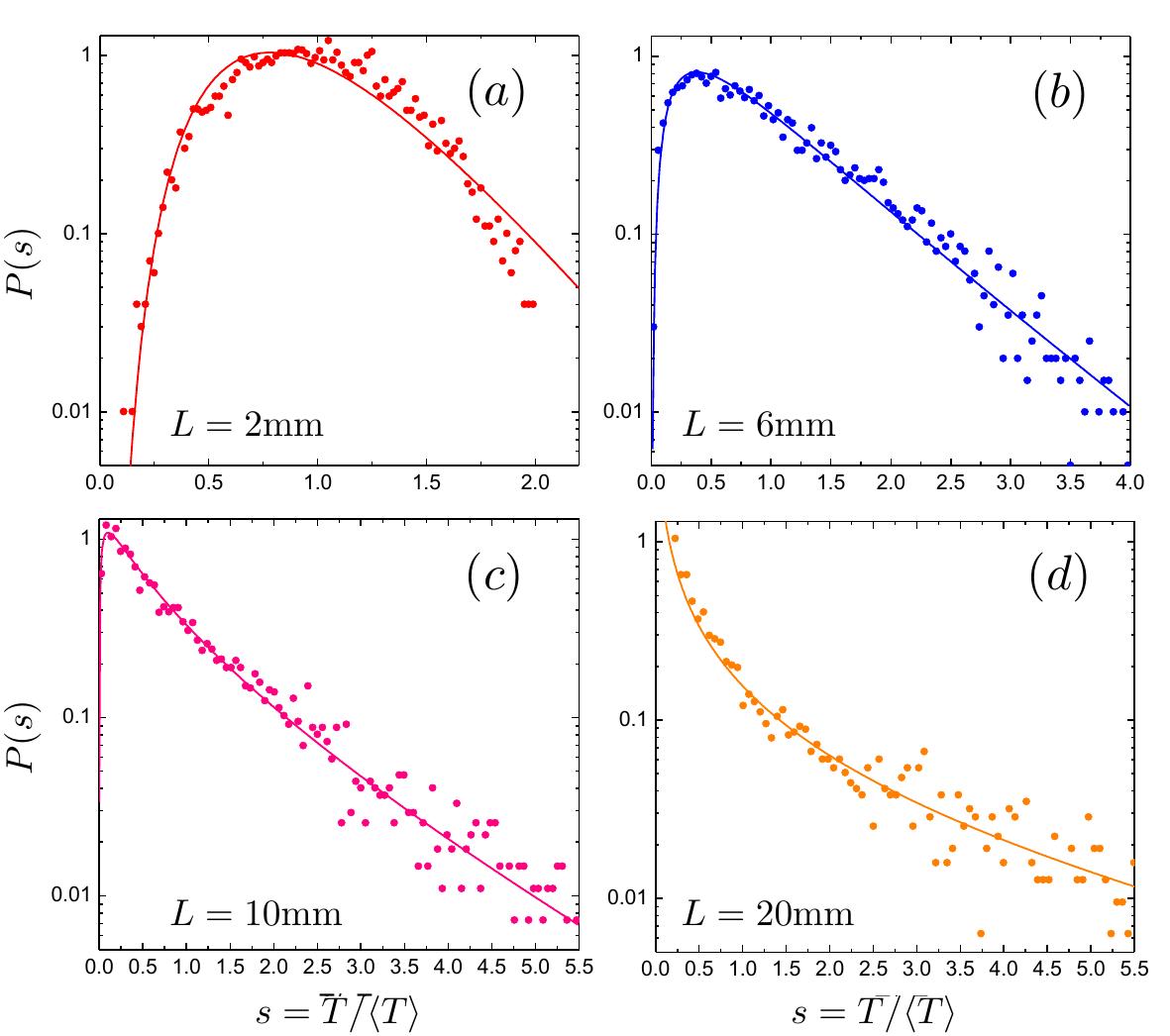}}
\caption{Selected probability distributions of normalized transmission $s = T/\langle T \rangle$ for $\phi = 0.079$, $f = 0.5$ THz, $n_{\mathrm{sc}} = 1$ and waveguide lengths $L = 2$ mm to $L = 20$ mm (a--d). The localization length is $\xi = (\pi/2) N l^* \simeq 9$ mm with $N=7$ and $l^* = 0.81$ mm [see Fig. \ref{Fig2}(a)].  Solid lines show fits to Eq.\ (\ref{ps1}) with the best-fit values of $g = \langle g \rangle$ shown by symbols in Fig.\ \ref{Fig5}(a).} \label{Fig4}
\end{figure}

\begin{figure}[t]
\centerline{\includegraphics[width=7.5 cm]{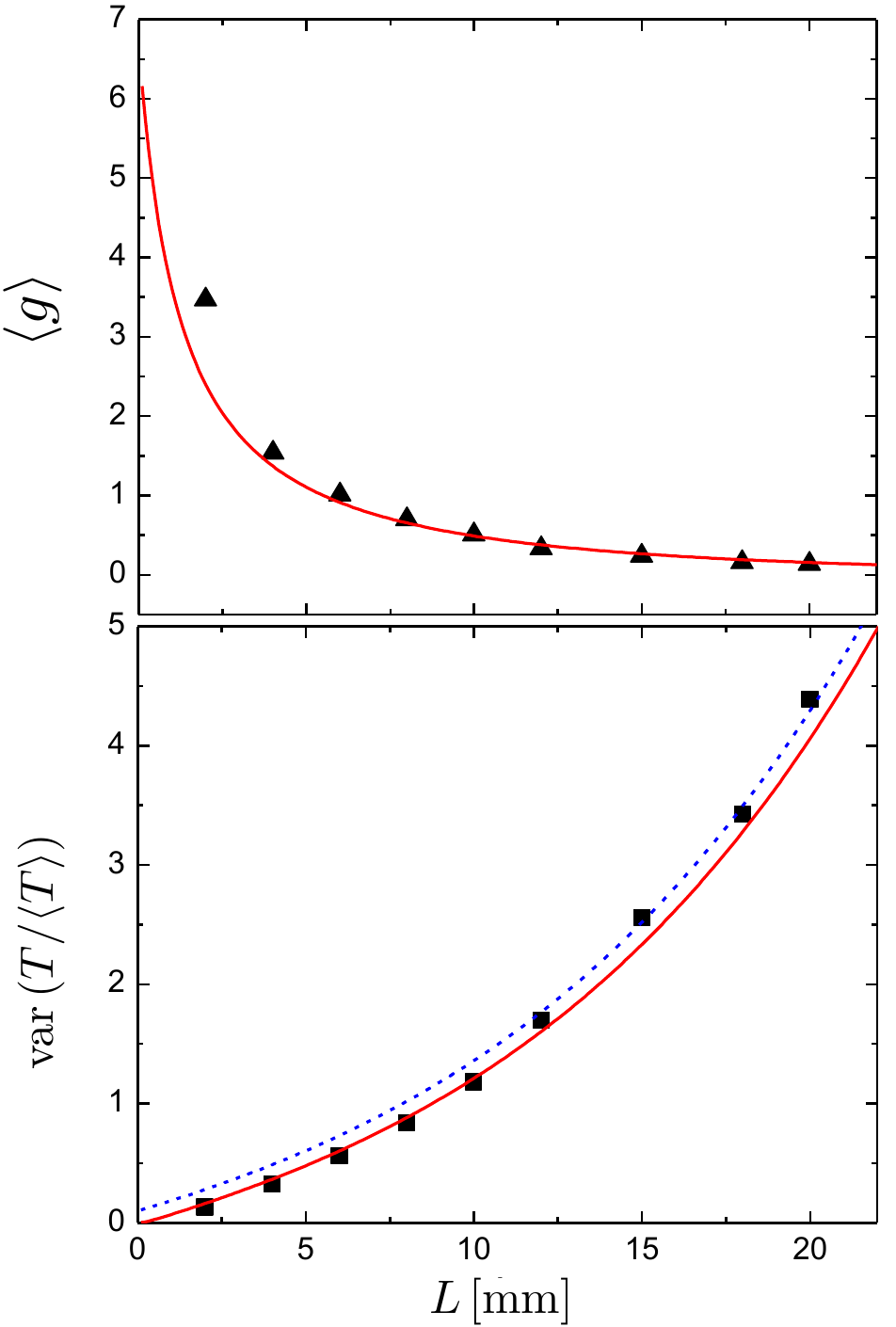}}
\caption{(a). Best-fit values of $\langle g \rangle$ found from the fits in Fig.\ \ref{Fig4} (symbols) compared to the theoretical prediction of Ref.\ \cite{MirlinPhysRep2000} (solid line) for $N=7$ and $l^* = 0.81$ mm. (b). Variance of normalized transmission as a function of waveguide length $L$ (symbols) compared to the theoretical results $\mathrm{var} (T/\langle T \rangle) = 2/3 \langle g \rangle$ \cite{vanRossumRevModPhys1999} (dotted line) and taking into account finite size effects \cite{SaenzFiniteSizePRL2002} (solid line).} \label{Fig5}
\end{figure}

Complementary to the average transmission, transmission fluctuations provide a second measure for the onset of localization \cite{MStoytchevPRL1997,AAChabanovNature2000,AZGenackJPhysAMathGen2005,SZhangSpeckleMicrowavePRL2007}. Using larger ensembles consisting of 5000 scatterer configurations, we determine probability distributions $P(s)$ of the normalized transmission $s = T/\langle T \rangle$ for $\phi = 0.079$, $N = 7$ and $f = 0.5$ THz and waveguide lengths ranging from $L = 2$ to 20 mm.
In Fig.\ \ref{Fig4}, numerical results are compared with the theoretical prediction \cite{Nieuwen1995,vanRossumRevModPhys1999}:
\begin{eqnarray}
P(s) &=& \int\limits_{-i \infty}^{i \infty} \frac{d x}{2 \pi i}
\mathrm{exp} \left[x s - \Phi(x) \right],
\label{ps1}
\\
\Phi(x) &=& g \ln^2 \left( \sqrt{1+x/g} + \sqrt{x/g} \right).
\label{ps2}
\end{eqnarray}
Theory predicts that the distribution is parameterized by a single parameter $g = \langle g \rangle$ that we adjust to fit the numerical data. As $L$ increases to values greater than $\xi$, $P(s)$ develops a tail that corresponds to a significant probability of obtaining large values of $T$ for certain realizations of disorder. The best-fit values of $\langle g \rangle$ are shown in Fig.\ \ref{Fig5}(a) (symbols) compared to the theoretical prediction following from Eq.\ (6.23) of Ref.\ \cite{MirlinPhysRep2000} where we replace the only parameter $\xi/L$ by $g_0$ given by our Eq.\ (\ref{dimcond}) and use $N = 7$ and $l^* =0.81$ mm obtained from the best fit to the average transmission data [see Fig.\ \ref{Fig2}(a)]. The agreement between the best-fit values of $\langle g \rangle$ extracted from the fits to the numerical data for $P(s)$ and the theory is good.

The large fluctuations of $T$ can be quantified by the variance of $s = T/\langle T \rangle$ that we show in Fig.\ \ref{Fig5}(b) as a function of waveguide length $L$. Numerical data (symbols) are compared with the theoretical results $\mathrm{var}(s) = 2/3 \langle g \rangle$ \cite{vanRossumRevModPhys1999, AAChabanovNature2000, AZGenackJPhysAMathGen2005} (dashed line)
and $\mathrm{var}(s) = (2/3 \langle g) \rangle
(1+3l^*/2L)/(1+l^*/L)^3$ \cite{SaenzFiniteSizePRL2002} (solid line), where $\langle g \rangle$ is obtained from Eq.\ (6.23) of Ref.\ \cite{MirlinPhysRep2000} [see the solid line in Fig.\ \ref{Fig5}(a)].
The equation derived in Ref.\ \cite{SaenzFiniteSizePRL2002} takes into account the finite-size effects and better agrees with the numerical results for short samples whereas the simpler equation $\mathrm{var}(s) = 2/3 \langle g \rangle$ seems to do a better job at large $L > 10$ mm. We would like to stress here that lines in Fig.\ \ref{Fig5} are not fits to numerical data but theoretical results obtained using $l^* = 0.81$ mm extracted from fits of Fig.\ \ref{Fig2}. Good overall agreement between theory and numerics, together with the good quality of fits shown in Fig.\ \ref{Fig4}, suggest that Eq.\ (\ref{ps1}) for the distribution of total transmission may be a good approximation even in the localized regime, provided that the parameter $g$ is understood as $\langle g \rangle$ and computed using the exact theory available for this quantity. This is a nontrivial result because Eq.\ (\ref{ps1}) was initially derived for weak disorder and can be rigourously justified only for $g = g_0 \gg 1$. Indications of qualitative validity of Eq.\ (\ref{ps1}) in the localized regime were already contained in some of the previous experimental studies \cite{MStoytchevPRL1997,AAChabanovNature2000,AcousticLocalization2008}. In contrast to these studies in which the actual values of $\langle g \rangle$ were not known, our results allow for a quantitative test of Eq.\ (\ref{ps1}) because we can compare the values of $\langle g \rangle$ extracted from the fits with the values following from the exact theory [see Fig.\ \ref{Fig5}(a)]. Such a comparison shows that Eq.\ (\ref{ps1}) has quite a reasonable degree of precision that is likely to be sufficient for description of experimental data.

\section{Conclusions}

Our numerical results clearly demonstrate the feasibility of observation of Anderson localization of THz waves in quasi-1D disordered waveguides under realistic experimental conditions. Due to the high contrast in refractive index between the scatterers and the matrix medium attainable in the THz frequency range, the localization length can be made quite short, typically of the order of several mm, already for moderate area filling fractions of scatterers $\phi \sim 0.1$. The results of our numerical simulations are in overall good quantitative agreement with existing approximate and exact theories for the average transmission \cite{MirlinPhysRep2000,PaynePRB2010} and  its full statistical distribution \cite{MStoytchevPRL1997,vanRossumRevModPhys1999} provided that in the latter case, the parameter $g$ of the distribution is understood as the average conductance $\langle g \rangle$ and calculated using the exact theory available for it (see, e.g., Ref.\ \cite{MirlinPhysRep2000}) and not as the bare conductance $g_0$ defined through the geometrical parameters of the waveguide according to Eq.\ (\ref{dimcond}). Moreover, our study shows that all relevant length scales, such as the scatterer size $a$, the waveguide width $w$ and length $L$, the transport mean free path $l^{*}$ are rather well separated which allows to clearly distinguish between different regimes of propagation. Finally, structures with design parameters close or even exactly equal to those used in our simulations can be manufactured with high precision as has been previously demonstrated \cite{PPeierTHzStrucWGs2010}. This opens the door for future experimental studies of EM wave localization with THz waves that promise improved spatial resolution and better experimental access to the EM field amplitude rather than intensity.

\acknowledgments

The present project has been financially supported by the the Swiss National Science Foundation (projects 132736, 142571, 140943 and 149867) and the Swiss State Secretariat for Education, Research and Innovation (SERI). S.E.S. acknowledges support from the Federal Program for Scientific and Scientific-Pedagogical Personnel of Innovative Russia for 2009--2013 (contract No. 14.B37.21.1938).

\end{document}